\documentclass[10pt,journal,compsoc]{IEEEtran}
\usepackage{cite}
\usepackage{amsmath,amssymb,amsfonts}
\usepackage{algorithmic}
\usepackage{booktabs}
\usepackage{graphicx}
\usepackage{textcomp}
\usepackage{adjustbox}
\usepackage{subcaption}
\usepackage{listings}
\usepackage{caption}
\usepackage{color}
\usepackage{multirow}
\usepackage{array}
\usepackage[bookmarks=false]{hyperref}
\def\fixme#1{\typeout{FIXED in page \thepage : {#1}}
\bgroup \color{red}{[FIXME: {#1}]} \egroup}

\def\BibTeX{{\rm B\kern-.05em{\sc i\kern-.025em b}\kern-.08em
    T\kern-.1667em\lower.7ex\hbox{E}\kern-.125emX}}

\begin{document}

\title{Memory-Aware Denial-of-Service Attacks on Shared Cache in
  Multicore Real-Time Systems\thanks{This research is supported in part by NSF CNS 1718880, CNS 1815959, and NSA Science of Security initiative contract \#H98230-18-D-0009.}} 

\pagestyle{empty}
\author{Michael Bechtel, Heechul Yun\\
  University of Kansas, USA. \\
  \{mbechtel, heechul.yun\}@ku.edu \\
}

\maketitle

\begin{abstract}
  In this paper, we identify that memory performance plays a crucial role in
  the feasibility and effectiveness for performing denial-of-service
  attacks on shared cache. Based on this insight, we introduce new
  cache DoS attacks, which can be mounted from the user-space and can
  cause extreme worst-case execution time (WCET) impacts to cross-core victims---even if the
  shared cache is partitioned---by taking advantage of the platform's
  memory address mapping information and HugePage support.
  We deploy these enhanced attacks on two popular embedded out-of-order 
  multicore platforms using both synthetic and real-world benchmarks.
  The proposed DoS attacks achieve up to 111X WCET increases on the tested
  platforms.
\end{abstract}


\section{Introduction} \label{sec:intro}

Multicore computing platforms are increasingly used in safety-critical
cyber-physical systems such as self-driving cars and drones. However,
in a multicore platform, a task's execution time can vary
significantly due to contention in shared micro-architectural
resources when other tasks run concurrently on the
platform~\cite{bosch2019challenge}. Such timing variation in multicore 
can be exploited by attackers.
Consider, for example, a scenario where some cores of a multicore
platform are reserved for critical real-time tasks while some other cores are
reserved for user downloaded third party programs. Even if the platform's
runtime (OS or hypervisor) partitions cores and memory to isolate the
potentially dangerous programs from the critical tasks, as long as they
share the same multicore computing platform, 
a malicious program may still delay the critical tasks by executing code 
that effectively mounts denial-of-service (DoS) attacks.


Modern multicore processors provide a high-degree of parallelism in
accessing memory throughout the memory hierarchy.
At the cache-level, non-blocking caches~\cite{kroft1981lockup} are
used, and can be accessed even when there are multiple outstanding
cache misses.
However, a non-blocking cache can become inaccessible whenever its
internal hardware buffers are exhausted, at which point the cache
is blocked and cannot accept any further requests. The cache then remains blocked until
the internal buffers become available
again~\cite{gem5memory,shen2013modern}.
For a shared last level cache (LLC), cache blocking is especially problematic because
it affects all cores that share the cache, as all requests to the cache would be 
blocked. 
As a result, the cores need to wait for the cache to unblock, which
can take a long time as the cache may need to access slower main 
memory, which can take hundreds of CPU cycles.
Therefore, if an attacker can intentionally induce cache blocking on
the shared LLC, they can cause massive timing impacts to the rest of the
cores even if they cannot directly access them. 


Prior work demonstrated the feasibility and severity of
micro-architectural DoS attacks~\cite{bechtel2019dos,bechtel2018picar,valsan2016taming} on
shared non-blocking caches, which identified two internal cache hardware
structures: (1) miss-status-holding-registers (MSHRs), which track
individual requests generated from cache misses, and (2) write-back
buffers, which temporarily hold and delay cache write-backs, as
potential DoS attack vectors.
In these works, an attacker simply accesses a large array and
quickly generates a large number of concurrent cache-misses. This then exhausts
the cache internal structures and effectively induces cache blocking.
They also showed that conventional cache partitioning
techniques are ineffective to defend against such DoS attacks that
target internal cache hardware structures because they can still be shared even if the cache space is partitioned.

In this paper, we propose \emph{memory-aware cache DoS attacks}, which
can induce more effective cache blocking by taking advantage of the 
memory address mapping information of the underlying memory hardware. 
Like prior cache DoS attacks, our new attacks also generate
continuous cache misses to exhaust shared cache internal hardware
resources. The difference is that we carefully control those cache
misses to target the same DRAM bank. Because concurrent accesses to the same DRAM bank cannot take advantage of bank-level parallelism and incur lots of DRAM bank conflicts, they will be slower to process~\cite{yun2015ecrts}, 
which in turn will lengthen the duration of cache blocking, helping to improve our attack. 
To realize this, we leverage Linux's HugePage support to directly control part of a 
physical address so as to control its DRAM bank location in allocating memory.


We implement and validate the proposed memory-aware DoS attacks on two contemporary
embedded multicore platforms---Raspberry Pi 4 and Odroid XU4---using
both synthetic and representative real-world benchmarks. We find that the proposed memory-aware cache DoS attacks are significantly and consistently more effective at impacting a victim task's execution time (observed up to 56X slowdown), compared to state-of-the-art cache DoS attacks. 

\section{Background} \label{sec:background}

In this section, we provide necessary background on non-blocking caches, main memory, and HugePage.

\subsection{Non-Blocking Cache}


Modern processors employ non-blocking caches, which employ multiple
internal hardware structures, such as Miss-Status-Holding-Registers (MSHRs) and the WriteBack (WB) buffer, to support parallelism in accessing memory~\cite{shen2013modern}.

On a non-blocking cache, when a cache-miss occurs, an MSHR entry is
allocated to record the miss related information.
The MSHR entry is then cleared only when the desired cache-line is
returned from the lower levels of the memory hierarchy (e.g., LLC,
DRAM). 
Multiple outstanding cache-misses can be
supported by a non-blocking cache, although the degree to which it can
happen depends on the size of the cache MSHR, which determines the
cache's memory-level parallelism (MLP). For the remainder of this paper,
we use the terms local MLP and global MLP as the number of MSHRs in a private
cache and a shared LLC, respectively.


On the other hand, the WriteBack buffer holds dirty cache-lines that
are evicted from the cache and need to be written back to the next level in
memory. Because reads from memory, such as the cache refills
generated from cache-line evictions, are generally more important for
application performance, delaying writebacks to memory while reads are
being processed can improve system performance by reducing bus
contention. The writebacks are then sent to memory when there are no
reads being serviced or when the buffer is full. In this way, a
non-blocking cache can support concurrent access to the cache
efficiently most of the time.

Note, however, that when either the MSHRs or WriteBack buffer
become full, the entire cache is blocked and  rejects all subsequent
requests until the cache is unblocked when both structures have free
entries available. Unfortunately, unblocking can take a relatively
long time as it depends on response times from the lower memory
levels. In the worst case, it can take upwards of hundreds of CPU
cycles when accesses to the slower main memory are required.
\emph{Cache blocking} is especially problematic in a shared cache as it
affects all cores that share that cache.
Even if a task's memory accesses are all cache hits, the task can 
still suffer massive slowdowns if the cache is blocked for a
significant portion of the time.

\subsection{Main Memory (DRAM)}
A DRAM chip is organized to have one or more ranks, each consists of
multiple banks~\cite{jacob2010memory}.
A bank contains storage cells, which are organized in rows and columns in a
2D array-like structure. To access data in the storage cells, the
corresponding row must be opened (activated), which copies the data of the row
into an intermediary buffer, called a row buffer, which acts as a cache. 
While in the row buffer, the data can be read from/written to efficiently. 
To access a different row, however, 
the current row needs to be closed (precharged). 
Since both activation and precharge take considerable time, accesses
to different rows in the same bank can decrease memory performance.

To access specific locations in the memory, the system's
DRAM controller employs a memory address mapping scheme that translates
a given physical address to the DRAM specific addresses (rank, bank, row, and column).
Because DRAM banks can be accessed in parallel, the address mapping of DRAM banks is 
particularly important for memory performance. If concurrent memory requests are mapped over
different banks, they can be processed efficiently in parallel and thus faster; if,
however, they are mapped to the same bank, resulting \emph{bank conflicts} will slowdown 
the memory performance~\cite{yun2014rtas}.
Therefore, if an attacker can control the physical addresses in allocating memory, 
they can also control which DRAM banks the allocated memory blocks will be located on. This, in turn, will cause a slowdown in memory performance by intentionally generating lots of bank conflicts.



\subsection{HugePage}
In a virtual memory-based system, memory is typically allocated in a 4KB
page granularity. However, most modern architectures, including ARM, support 
bigger page sizes (e.g., 2MB pages) and Linux's \textit{HugePage}~\cite{hugepage} 
infrastructure gives applications an option to use them when allocating memory. 
This can reduce the number of pages used by applications and the 
CPU's translation look-aside buffer (TLB) pressure~\cite{panwar2018making}, as each TLB 
entry can cover a larger address range (2MB vs. 4KB). This 
can improve performance and predictability, 
which are needed for many modern real-time systems.
As such, many recent embedded platforms and their Linux kernels support HugePages.


\begin{figure}[h]
  \centering
  \includegraphics[width=.40\textwidth]{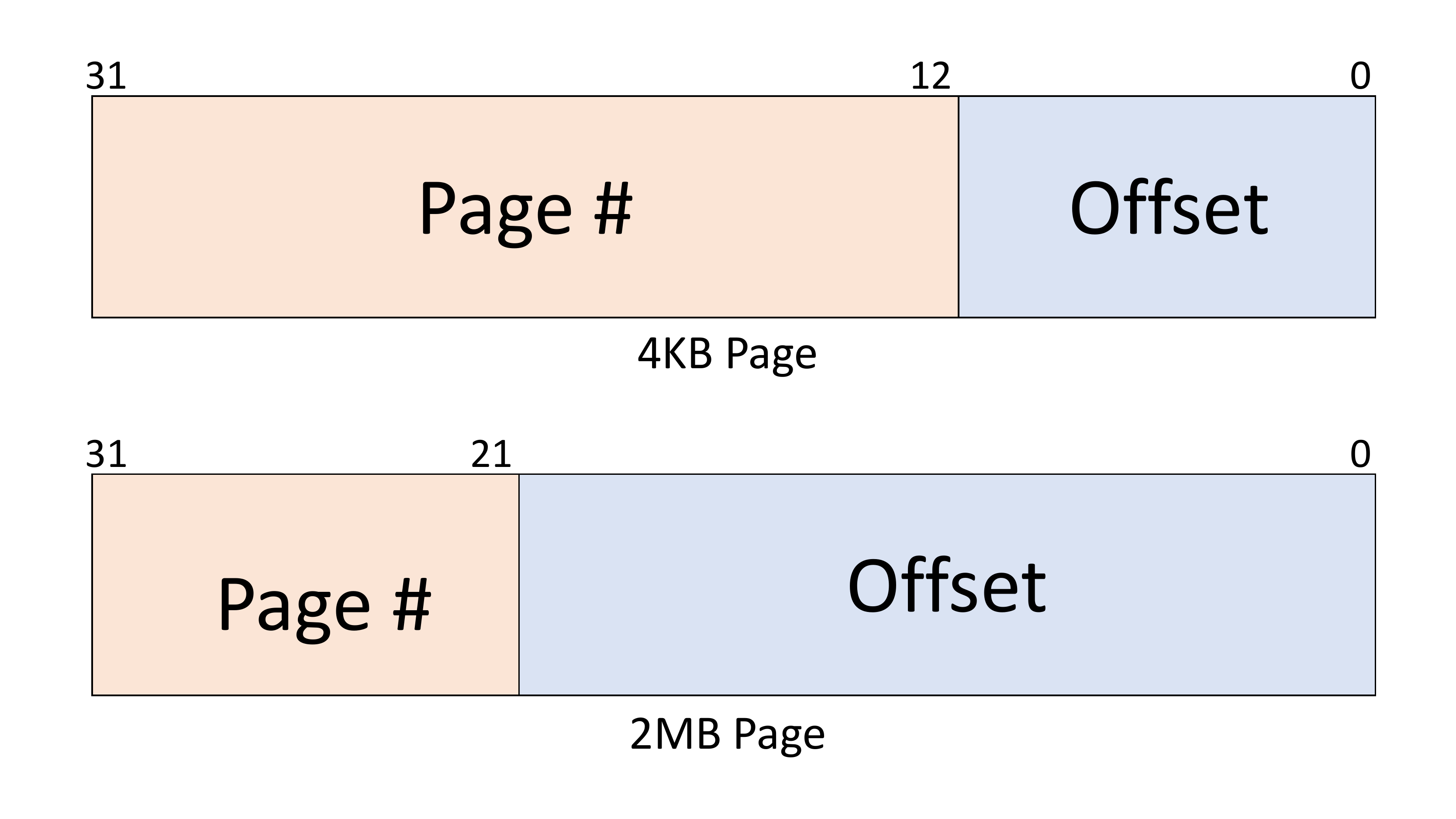}
  \caption{Virtual address mappings in 4KB and 2MB pages.}
  \label{fig:address-mapping}
\end{figure}

Figure~\ref{fig:address-mapping} shows the virtual address mappings
for 4KB and 2MB page granularity on a 32-bit system. For the 4KB
granularity, 12 bits are used in the offset of a page, while 21 bits are
used for the offset of a 2MB page. 
Note that this means that allocating a single 2MB page allows us to
control a larger portion of the physical address space (the lowest 21 bits) without requiring
any system privileges.
In the following, we will exploit this ability to control (part of) physical addresses 
for creating effective cache DoS attacks. 


\section{Threat Model}\label{sec:model}



We assume that victim and attacker are co-located on a multicore
processor which features multiple processing cores, per-core private caches, 
and a single shared last-level cache (LLC).
We consider a runtime system (OS and hypervisor) that 
can partition core, memory, and LLC space to limit resource
sharing between the victim and the attacker.
In addition, the attacker's capability is limited to executing non-privileged user-level code.
The aforementioned assumptions are the
same as those used in the prior work~\cite{bechtel2019dos}.

In this work, we make the following two additional assumptions:
First, the runtime supports HugePages and the attacker can use it to allocate memory in its own private  address space.
Second, the attacker knows the system's physical address to DRAM bank address mapping information. 
Note that major architectures (e.g. x86, ARM) and OSes (e.g. Linux) support HugePages and prior works showed that it is possible to reverse engineer the DRAM bank mapping information on a variety of computing platforms~\cite{helm2020reliable,pessl2016drama,liu2012software}.
As such, we believe these additional assumptions are realistic. 
We provide further discussion on these assumptions in Section~\ref{sec:discussion}. 

In this setting, the attacker's primary goal is to delay the execution
time of the victim task by mounting denial-of-service attacks on the
shared cache.

\section{Memory-Aware Cache DoS Attack}~\label{sec:attackers}

In this section, we discuss memory access characteristics of
prior cache DoS attacks and their limitations (Section~\ref{sec:seqattack}), followed by the proposed
memory-aware cache DoS attacks (Section~\ref{sec:plattack} and \ref{sec:attackcode}).

\subsection{Sequential Attack}~\label{sec:seqattack}


\begin{figure}[h]
  \centering
  \captionsetup[subfigure]{oneside,margin={1.2cm,0cm}}
  \begin{subfigure}[b]{0.4\columnwidth}
\begin{lstlisting}[linewidth=\textwidth,language=c,xrightmargin=-.18\textwidth]
for (i = 0; i<mem_size; 
	i += LINE_SIZE)
{
	sum += ptr[i];
}
\end{lstlisting}
    \caption{BwRead}
    \label{fig:read}
  \end{subfigure} \hfill
  \begin{subfigure}[b]{0.4\columnwidth}
\begin{lstlisting}[linewidth=\textwidth,language=c,xrightmargin=-.18\textwidth]
for (i = 0; i<mem_size; 
	i += LINE_SIZE)
{
	ptr[i] = 0xff;
}
\end{lstlisting}
    \caption{BwWrite}
    \label{fig:write}
  \end{subfigure}
  \caption{Sequential memory access attacks. }
  \label{fig:bwattackcode}
\end{figure}

Figure~\ref{fig:bwattackcode} shows the code snippets of the prior
cache DoS attacks~\cite{valsan2016taming,bechtel2019dos}, which
perform a series of sequential memory accesses over a large array.

The \emph{BwRead} attack iteratively reads entries of a large
one-dimensional array at a cache-line granularity (LINE\_SIZE,
which is typically 64 bytes). When \textit{mem\_size} is larger than the size
of the LLC, it generates lots of cache-misses and, in turn, accesses to
main memory. On a modern processor, multiple cache-misses can occur
concurrently---with the help of out-of-order execution and/or hardware
prefetchers---which may stress the LLC's MSHRs~\cite{valsan2016taming,bechtel2019dos}.

The \emph{BwWrite} attack operates in a similar manner, but instead writes a
value to each array entry. This will then generate continuous store
operations that can also be configured to intentionally miss the
LLC. Again, multiple write misses can occur concurrently, which can
stress both the MSHRs and the Writeback Buffer of the cache. This is
because each missed write can generate up to two memory requests: a
read for a cache linefill and a write for a cache
writeback~\cite{bechtel2019dos}.

While these attacks are effective at generating a large number of
concurrent cache misses,
their sequential memory access nature 
means these misses can be processed efficiently at the memory level.
Concretely, successive cache misses are likely to be
allocated on the same DRAM row (e.g. 2KB in LPDDR4) and thus are processed efficiently 
at the DRAM because costly row switching is not needed.

Note that efficient processing at memory is \emph{undesirable} from the
perspective of a cache DoS attack because its goal is to
induce longer cache blocking and fast memory performance would instead reduce
the duration of cache blocking.


\subsection{Parallel Linked-List Attack}\label{sec:plattack}
 
To address the shortcomings of the sequential memory access-based
cache DoS attacks, we first introduce parallel linked-list attacks,
which generate concurrent random memory accesses.

\begin{figure}[h]
  \centering
  \captionsetup[subfigure]{oneside,margin={1.2cm,0cm}}
  \begin{subfigure}[b]{0.4\columnwidth}
\begin{lstlisting}[linewidth=\textwidth,language=c,xrightmargin=-.18\textwidth]
static int* list[MAX_MLP];
static int next[MAX_MLP];

for (int64_t i = 0; i < iter; i++) {
    switch (mlp) {
    case MAX_MLP:
    .
    .
    case 2:
        next[1] = 
			list[1][next[1]];
        /* fall-through */
    case 1:
        next[0] = 
			list[0][next[0]];
    }
}
\end{lstlisting}
z    \caption{PLLRead}
    \label{fig:read}
  \end{subfigure} \hfill
  \begin{subfigure}[b]{0.4\columnwidth}
\begin{lstlisting}[linewidth=\textwidth,language=c,xrightmargin=-.18\textwidth]
static int* list[MAX_MLP];
static int next[MAX_MLP];

for (int64_t i = 0; i < iter; i++) {
    switch (mlp) {
    case MAX_MLP:
    .
    .
    case 2:
        list[1][next[1]+1] = 
			0xff;
        next[1] = 
			list[1][next[1]];
        /* fall-through */
    case 1:
        list[0][next[0]+1] = 
			0xff;
        next[0] = 
			list[0][next[0]];
    }
}
\end{lstlisting}
    \caption{PLLWrite}
    \label{fig:write}
  \end{subfigure}
  \caption{Parallel linked-lists attacks. 
    Linked-list entries are randomly shuffled over
    a large address space.}
  \label{fig:lmlpattackcode}
\end{figure}

Figure~\ref{fig:lmlpattackcode} shows the code snippets for the
parallel linked-list attacks: \emph{PLLRead} for read and \emph{PLLWrite}
for write. In both cases, the attacks traverse a set number of linked
lists, which can be accessed concurrently on a modern \emph{out-of-order}
core because there is no data dependency between the entries of
different lists. Each linked list is randomly shuffled over a
large memory address space to prevent prefetching.
As such, the number of linked lists determines the degree of 
memory-level parallelism (MLP) of the attacks. 
Note that the parallel-linked list attacks are based
on the MLP measurement code in~\cite{david2012bandit}.




Like the sequential access attacks, the parallel-linked list attacks
are designed to generate concurrent cache-misses, which would stress 
cache internal hardware buffers and induce cache blocking.
Unlike the sequential attacks, though, they are potentially less efficient in memory---hence more 
effective DoS attacks---because memory requests from a linked-list are likely mapped on different DRAM rows, which would 
require costly row switching~\cite{yun2015ecrts}.
Note, however, that entries of different linked-lists can still be mapped to
different DRAM banks. This would mean that concurrent accesses to different lists can be 
processed in parallel on different banks, which can hide the increased overhead 
of frequent row-switching at each individual bank.
This is undesirable from the perspective of a 
cache DoS attack because it needs slower---not faster---memory performance to 
be more effective. 



\subsection{DRAM Bank-Aware Parallel Linked-List
  Attack}~\label{sec:attackcode}

To overcome the limitations of the aforementioned cache DoS attacks, 
we propose a DRAM bank-aware cache DoS attack,
which is based on the parallel-linked list attack code
(Section~\ref{sec:plattack}) but differs 
in that the entries of the linked-lists are constructed in such a way
that they are all allocated in the \emph{same} DRAM bank. The rational is
that when multiple accesses target the same bank, they will take
longer to be serviced at the DRAM because of increased DRAM bank
conflicts and frequent row switching.



\begin{figure}[h]
  \centering
  \begin{lstlisting}[linewidth=0.45\textwidth,language=c]
int paddr_to_bank(unsigned long mask, unsigned long addr)
{
	int bank = 0;
	int idx = 0;
	int bit;
	for_each_set_bit(bit, &mask, BITS_PER_LONG) {
		if ((addr >> (bit)) & 0x1)
			bank |= (1<<idx);
		idx++;
	}
	return bank;
}
\end{lstlisting}

  \caption{Physical address to DRAM bank mapping function.}
  \label{fig:att2list}
\end{figure}

Figure~\ref{fig:att2list} shows the physical address to DRAM bank mapping function we used in this work. 
It checks the value of each bit in a given address that corresponds to
the set of bits specified in the \textit{mask} bitmask, which is the
platform's physical address bits that are mapped to DRAM banks.
By comparing the set bits in \textit{mask} with that of the given address \textit{addr}, 
we can determine which DRAM bank the address belongs to. 
Note that depending on the hardware platform, a more complex mapping function may be needed 
(e.g. multiple rounds of XOR operations:~\cite{helm2020reliable,pessl2016drama}). 

We use the mapping function in allocating memory blocks for the entries in the linked-lists as follows. 
First, we allocate a big chunk of memory using Linux's HugePage support (i.e. 2MB pages). 
Then, when constructing a linked list, we randomly select an address within the big chunk. 
If the candidate address's bank index (return value of the \textit{paddr\_to\_bank()} function) is zero, 
we add the address
to the linked list as a new entry, otherwise we discard the address and continue with a 
different randomly picked address, until we construct all entries with the same bank number. 

Because 2MB pages are used for memory allocation, up to 21 bits of a virtual address are the same as its corresponding 
physical address. This can effectively allow the attacker control a large number of physical address bits, including those that determine the DRAM bank allocation. 
As a result, when all of the linked lists are generated, all of their entries will be allocated on the same memory bank. When these linked lists are accessed concurrently by 
an out-of-order core, all the memory requests will then target the same DRAM bank. Furthermore, they also likely target 
different rows of the bank because of random addressing. As a result, their memory performance will 
be very slow, which in turn result in longer cache blocking and, in turn, more effective cache DoS attacks.

\section{Evaluation}\label{sec:evaluation}

In this section, we evaluate the effectiveness of the proposed
memory-aware cache DoS attacks on two embedded multicore-based
platforms using both synthetic and real-word applications.

\subsection{Embedded Multicore Platforms}~\label{sec:eval-platforms}

\begin{table}[h]
  \centering
  \begin{adjustbox}{width=.45\textwidth}
  \begin{tabular}{|c||c|c|}
    \hline
    Platform                & Odroid-XU4          & Raspberry Pi 4 (B) \\ \hline 
    SoC                     & Exynos5422          & BCM2711                \\ \hline 
    \multirow{3}{*}{CPU}    & 4x Cortex-A15 & 4x Cortex-A72          \\ 
                            & out-of-order  & out-of-order           \\ 
                            & 2.0GHz        & 1.5GHz                 \\ \hline 
    L1 (Private) Cache           & 32KB(I)/32KB(D)       & 48KB(I)/32KB(D)                \\ 
    L2 (Shared) Cache            & 2MB (16-way)  & 1MB (16-way)         \\ \hline 
    Memory                  & 2GB LPDDR3          & 4GB LPDDR4             \\ 
    (Peak BW)               & (14.9GB/s)          & (25.6 GB/s)            \\ 
    \hline
    DRAM Bank Bits               & 8, 13, 14, 15, 16   & 11, 12, 13, 14    \\ 
    \hline
  \end{tabular}
  \end{adjustbox}
  \caption{Compared embedded multicore platforms.}
  \label{tbl:platforms}
\end{table}

We deploy our DoS attacks on 
two embedded multicore platforms: an Odroid-XU4 and a Raspberry Pi 4 Model B.
The Odroid XU4 includes four Cortex-A7 (in-order) cores and four Cortex-A15 (out-of-order) cores, of which we only 
use the latter.
The second platform we test, the Raspberry Pi 4, only equips four Cortex-A72 (out-of-order) cores.
We reverse engineer the DRAM bank mapping information of the platforms using the methods described in~\cite{pessl2016drama,yun2014rtas}.
Table~\ref{tbl:platforms} shows the basic characteristics of the tested platforms. Note that the L2 cache is the last-level cache (LLC) in both platforms. 
As for the operating system, the Odroid-XU4 runs Ubuntu 18.04 and Linux kernel 4.14, while the Raspberry Pi 4 runs 
Raspbian Buster and Linux kernel 4.19.

\subsection{Impact to Synthetic Workloads}~\label{sec:eval-synthetic}

\begin{figure}[h]
  \centering
  \begin{subfigure}{0.45\textwidth}
    \includegraphics[width=\textwidth]{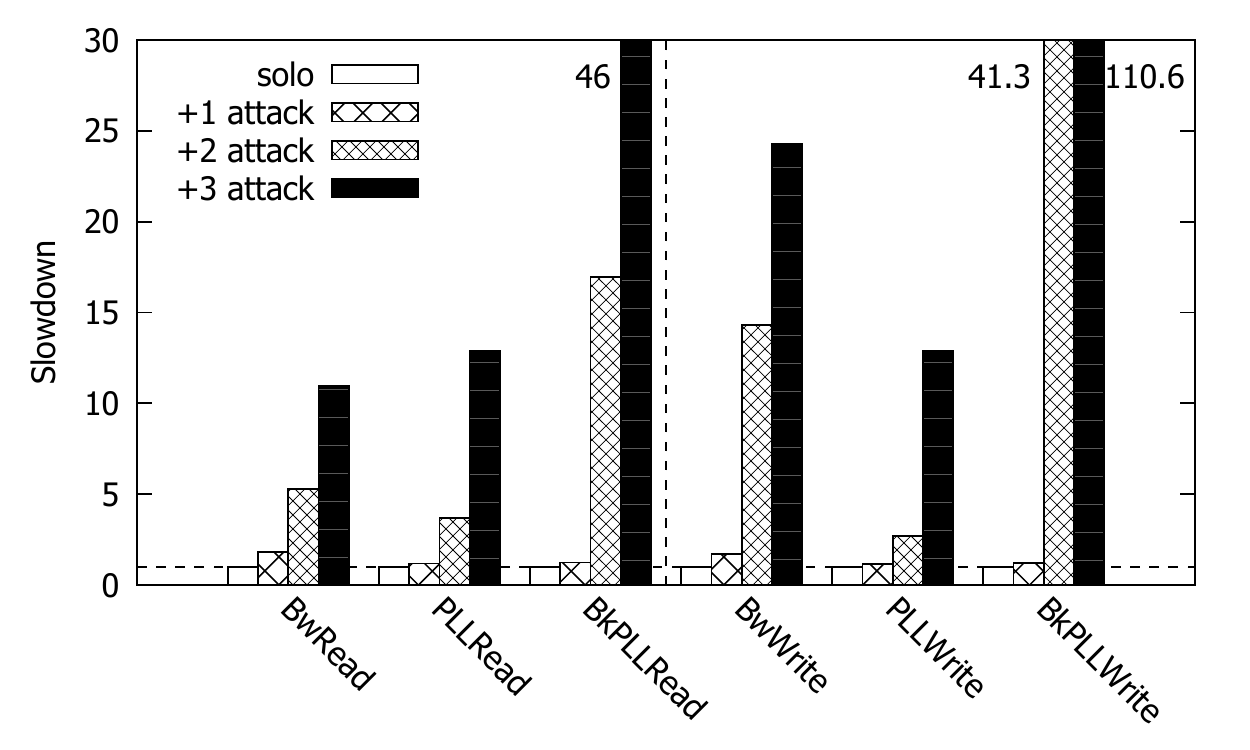}
    \caption{ Odroid XU4 (Cortex-A15)}
    \label{fig:xu4-a15-bwread}
  \end{subfigure}
  \begin{subfigure}{0.45\textwidth}
    \includegraphics[width=\textwidth]{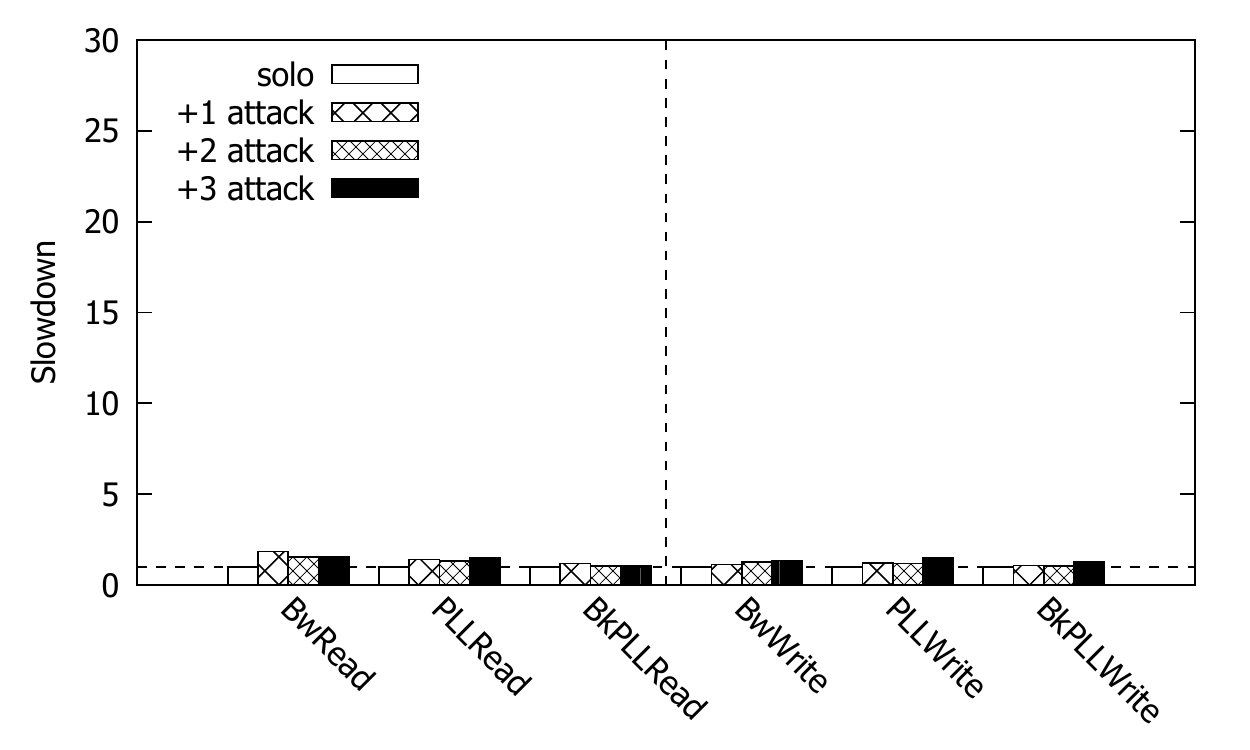}
    \caption{ Raspberry Pi 4 (Cortex-A72)}
    \label{fig:pi4-bwread}
  \end{subfigure}
  \caption{Effects of cache DoS attacks (X-axis) to a LLC fitting \emph{BwRead(LLC)} victim.}
  \label{fig:bwread-victim}
\end{figure}

\begin{figure}[h]
  \centering
  \begin{subfigure}{0.45\textwidth}
    \includegraphics[width=\textwidth]{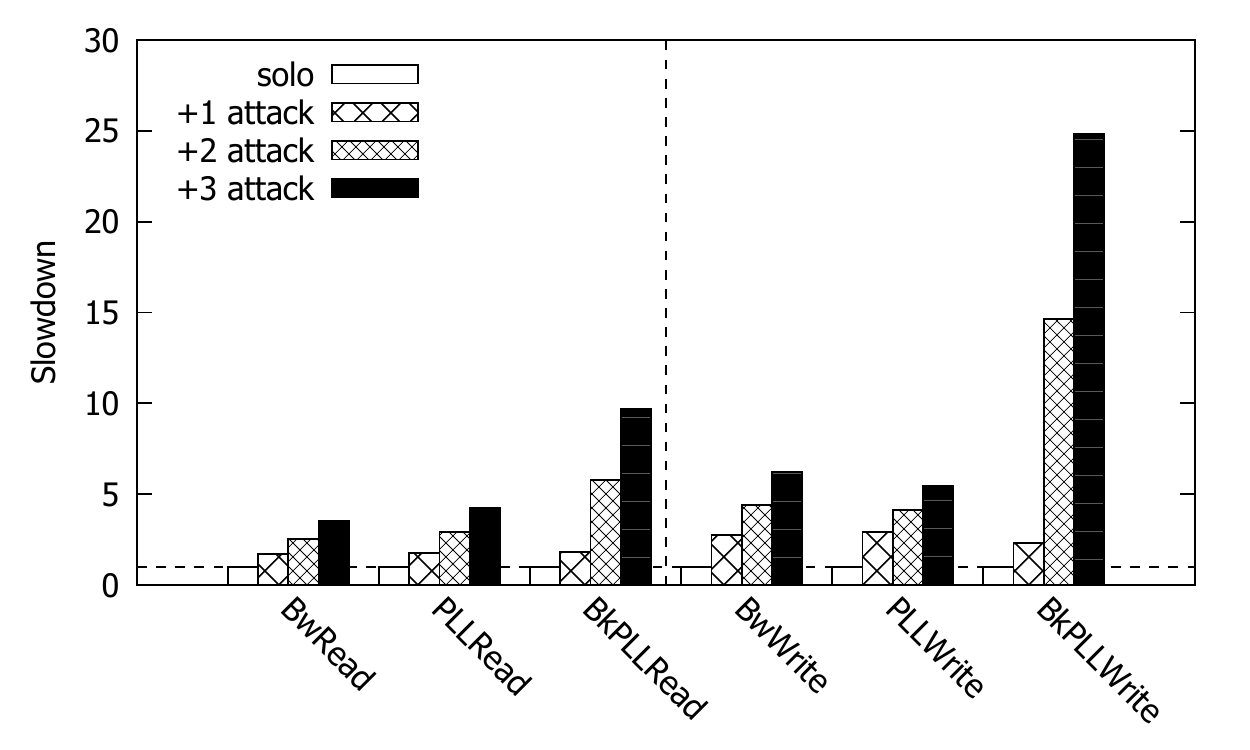}
    \caption{ Odroid XU4 (Cortex-A15) }
    \label{fig:xu4-a15-lmlp}
  \end{subfigure}
  \begin{subfigure}{0.45\textwidth}
    \includegraphics[width=\textwidth]{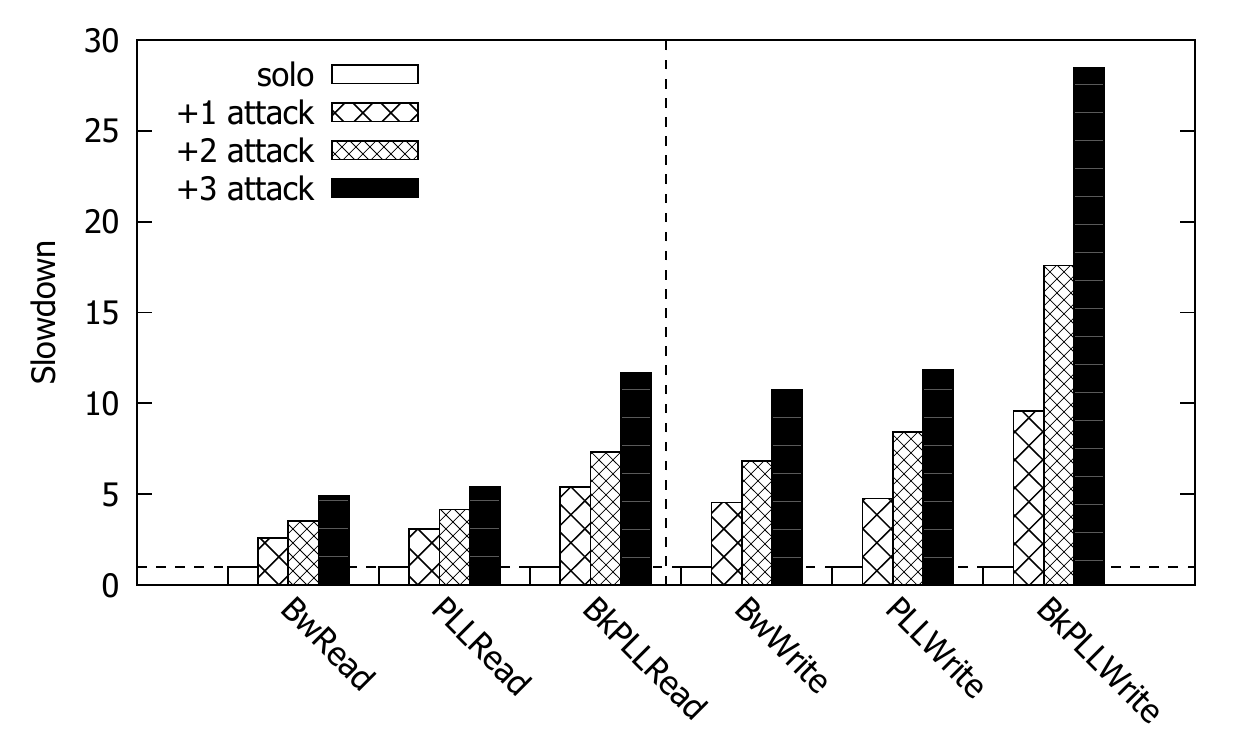}
    \caption{ Raspberry Pi 4 (Cortex-A72) }
    \label{fig:pi4-a72-lmlp}
  \end{subfigure}
  \caption{Effects of cache DoS attacks (X-axis) to a DRAM fitting \emph{BwRead(DRAM)} victim.}
  \label{fig:lmlp-victim}
\end{figure}

In this experiment, we evaluate the effectiveness of our memory-aware cache DoS attacks using synthetic workloads. 
The experimental setup is as follows: we first run a (synthetic) victim task alone
on a single core, Core 0, to measure its solo response time. We
then run the victim task alongside up to three instances of an attacker task, scheduled on Cores 1-3, and measure the response times of the victim task to determine the slowdown each attack caused on the victim relative to
the solo case.


For the victim tasks, we use BwRead(LLC) and BwRead(DRAM), both of which are based on the sequential DoS attack task (Section~\ref{sec:seqattack}) but differ in their working-set sizes---i.e. (LLC) means that the working-set is smaller than the platform's L2 cache size (but bigger than the L1 cache), whereas (DRAM) means that it is bigger than the L2 size.

For the attackers, we employ all three cache DoS attack types discussed in
Section~\ref{sec:attackers}, with each one capable of being read
intensive or write intensive, for a total of six attacking tasks. \emph{BwRead} and \emph{BwWrite} refer to read and write version, respectively, of the sequential memory attack in Section~\ref{sec:seqattack};  \emph{PLLRead} and \emph{PLLWrite} refer to the parallel linked-list attacks in Section~\ref{sec:plattack}; and \emph{BkPLLRead} and  \emph{BkPLLWrite} refer to the memory-aware (DRAM bank-aware) parallel 
linked list attacks in Section~\ref{sec:attackcode}.
For all attacking tasks, we configure their working set sizes to be bigger than the platform's L2 cache size. In other words, the attackers' working-set sizes are always (DRAM). As such, we drop the parenthesis when referring the attackers.

Figure~\ref{fig:xu4-a15-bwread} shows the effects of the cache DoS attacks to
the \emph{BwRead(LLC)} victim on Odroid-XU4. First, the sequential memory attackers---BwRead and BwWrite---are already quite effective on the Odroid-XU4, as they slowdown the cache fitting victim task more than 10 and 20 times, respectively. The results are consistent with the findings in the prior works~\cite{valsan2016taming,bechtel2019dos}, which 
suggested that the smaller number of LLC MSHRs in the Odroid-XU4's Cortex-A15 processor was the main culprit, and led to frequent cache blocking. 
Next, the parallel linked-list attackers---PLLRead and PLLWrite---show mixed results as PLLRead is slightly more effective than BwRead while PLLWrite is somewhat worse than BwWrite.
Lastly, our memory-aware parallel linked-list attackers---BkPLLRead and BkPLLWrite---are shown to be much more effective than the rest. The worst case slowdown of the victim was $\sim$46X when paired with three BkPLLRead attackers, and $\sim$111X with three BkPLLWrite attackers. This is because the outstanding cache-misses of the parallel linked lists cannot be processed efficiently in DRAM as they cannot leverage DRAM bank-level parallelism (all target a single bank) and most of them cause costly DRAM row switching (due to random access patterns). As a result, the attacks can generate more effective prolonged cache blocking, which slows down the victim as it frequently needs to access the blocked cache. 

On the other hand, Figure~\ref{fig:pi4-bwread} shows the results of the same experiment on the Raspberry Pi 4. Note that, unlike the Odroid-XU4, none of the attackers show significant performance impacts on the victim task. 
This can be explained by the fact that the Raspberry Pi4's Cortex-A72 features a significantly improved L2 cache  that can handle many more outstanding requests than that of Odroid-XU4's Cortex-A15. Concretely, the Cortex-A72's L2 cache can support up to 19 outstanding reads (or more depending on the implementation)~\cite{arm-cortex-a72}, while the Cortex-A15's L2 can handle only up to 11 outstanding reads~\cite{arm-cortex-a15, valsan2016taming}. 
As a result, it appears that all attackers---on their own---are unable to induce sufficient cache blocking
necessary to delay the BwRead(LLC) victim, which mainly accesses the L2 cache. 


In the next experiment, we instead use \emph{BwRead(DRAM)} as the victim task, which itself generates lots of L2 cache-misses.
Figure~\ref{fig:lmlp-victim} shows the results. First, note that on both platforms, our memory-aware attackers (BkPLLRead and BkPLLWrite) are significantly more effective than the rest, causing up to 28X slowdown on Raspberry Pi 4 and up to 25X on Odroid-XU4. While precise attributions are challenging due to the presence of various complex and opportunistic performance enhancing mechanisms (e.g., hardware prefetchers), the fact that our memory-aware attacks, which consume much less memory bandwidth by limiting its memory accesses to a single DRAM bank, caused significantly more slowdowns to the victim than other more bandwidth intensive attackers (both sequential and memory-unaware parallel linked-list attacks) suggest that the cause of the slowdown is due to increased cache blocking rather than DRAM bandwidth limitation. The results were similar when we explicitly partitioned DRAM banks between the victim and the attackers, further indicating that the observed slowdowns are due to cache blocking rather than DRAM related issues such as bandwidth or bank conflicts between the victim and the attackers. 



\begin{figure*}[h]
  \centering
  \begin{subfigure}{\textwidth}
    \centering
    \includegraphics[height=0.4\textwidth,width=0.9\textwidth]{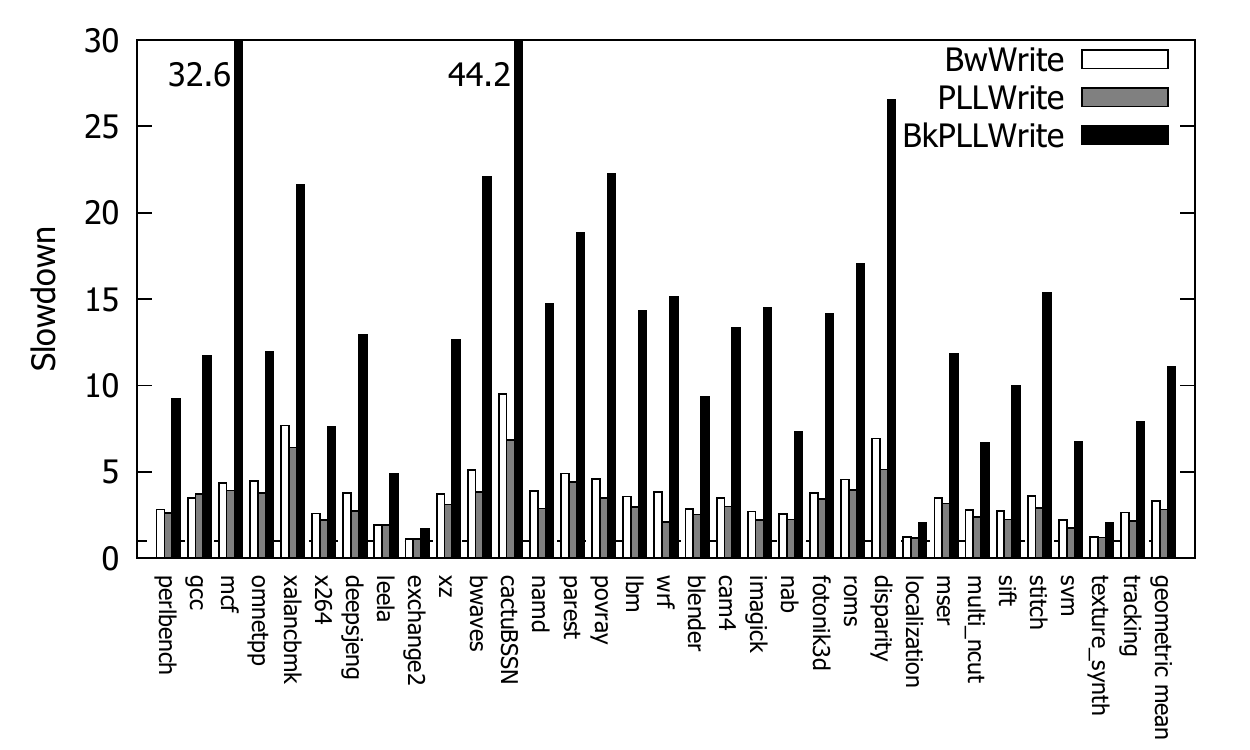}
    \caption{ Odroid XU4 (Cortex-A15) }
    \label{fig:realworld-xu4-write}
  \end{subfigure}
  \begin{subfigure}{\textwidth}
    \centering
    \includegraphics[height=0.4\textwidth,width=0.9\textwidth]{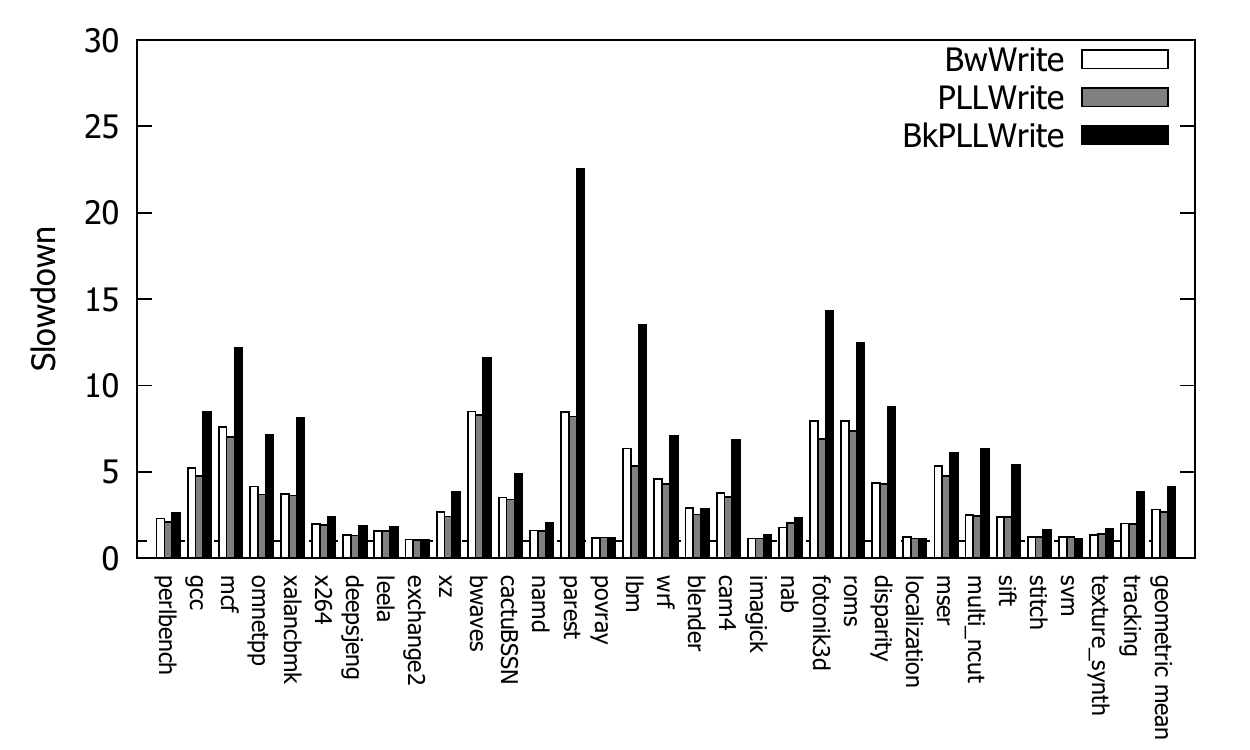}
    \caption{ Raspberry Pi 4 (Cortex-A72) }
    \label{fig:realworld-pi4-write}
  \end{subfigure}
  \caption{Effects of cache DoS attacks on SPEC2017 and SD-VBS benchmarks. 
    } 
  \label{fig:realworld}
\end{figure*}

\subsection{Impact to Real-World Applications}

In this experiment, we evaluate the effectiveness of our memory-aware cache DoS attacks using 32 real-world benchmarks from SPEC2017~\cite{spec2017} and SD-VBS~\cite{venkata2009sd}.
For each benchmark, we employ the same experimental methodology used
in Section~\ref{sec:eval-synthetic}. That is, we measure the victim's
execution time first alone in isolation and then together with three
instances of an attacker task. 
For the attackers, we only use the write versions (BwWrite, PLLWrite, and BkPLLWrite) as they were able to generate more contention than 
their respective read versions.

Figure~\ref{fig:realworld} shows the results.
On the Odroid-XU4, our memory-aware attack, BkPLLWrite, is able to delay the execution times of the 
real-world victim tasks far more effectively than other attackers, achieving a geometric mean of 11X slowdown (up to 44.2X for \textit{cactuBSSN}), which is 3.3X and 3.9X better than BwWrite and PLLWrite attacks, respectively. 
On the Raspberry Pi 4, on the other hand, BkPLLWrite achieves a geometric mean of 4.1X slowdown 
(up to 22.5X for \textit{parest}), which is 46\% and 52\% 
better than the BwWrite and PLLWrite attacks, respectively.
Interestingly, the improvements of BkPLLWrite are more pronounced in some benchmarks (e.g., \emph{parest}) while not significant on other benchmarks (e.g., svm) on the Raspberry Pi 4, whereas the improvements are significant in most of the tested benchmarks on Odroid-XU4. This is because those that primarily access the caches but not DRAM---similar to the BwRead(LLC) victim in Figure~\ref{fig:pi4-bwread}---would be less impacted by our attackers on Raspberry Pi 4 due to the reasons described in~\ref{sec:eval-synthetic}. 

In summary, we find that proposed memory-aware cache DoS attacks are substantially more effective than prior cache DoS attacks in increasing the execution times of real-world applications on both of the tested multicore platforms. 

\section{Discussion}~\label{sec:discussion}


In this section, we discuss limitations and possible future extensions of our work.

One notable shortcoming of the proposed memory-aware cache DoS attacks is that 
they do not work on in-order pipeline based processors (e.g. Cortex-A53). 
This is because, unlike an out-of-order core, an in-order core cannot traverse 
multiple linked lists concurrently and is thus unable to generate concurrent 
cache-misses and, in turn, induce cache blocking.
Note, however, that this does not mean successful cache DoS attacks are fundamentally 
impossible on in-order processors. In fact, prior work~\cite{bechtel2019dos} showed that how 
hardware prefetchers in in-order cores can be exploited 
to mount successful cache DoS attacks. Unfortunately, such hardware prefetchers
cannot follow the multiple linked lists used in our attacks. Memory-aware cache DoS attacks 
for in-order processors are left as future work. 


Another limitation of our work is that we assume that the attacker knows the 
memory address mapping scheme of the system. Depending on the hardware platform, this information can be difficult to obtain. First, most vendors do not publicly 
disclose the detailed memory mapping information. Second, reverse engineering can be compilcated 
if the memory controller employs a complex addressing scheme (e.g.~\cite{zhang2000permutation}), 
although there are sophisticated reverse engineering techniques (e.g.~\cite{helm2020reliable,pessl2016drama}) 
that can recover such complex mapping information. 
Lastly, a reverse engineering technique might require higher system-level privileges and 
additional information (e.g. prior knowledge on the number of DRAM banks) to be effective.
Note, however, an attacker does not need to perform the reverse engineering on the target 
platform that they intends to attack. Instead, it can be performed on any platform as long as 
it has the same processor and memory configuration because the mapping information would be the same.
While reliable and accurate reverse engineering of DRAM mapping information is still an important challenge, 
it is orthogonal to our work.

Third, our attacks assume that HugePage support is enabled and
can be used by the attacker in allocating its memory. Without HugePage support, only a 4KB
address space can be controlled by the attacker, which
is insufficient to control DRAM bank allocation on most
platforms. Therefore, disabling HugePages or only making them accessible to privileged users can also defeat our memory-aware DoS attacks.
Note, however, that HugePage support is common in most desktop and server platforms due to its potential performance benefits for large applications~\cite{panwar2018making}.
We also observe that increasingly many embedded platforms (e.g. NVIDIA's Jetson series)
include HugePage support by default to better support increasingly bigger and complex 
applications, especially those in intelligent robots.



Lastly, in this work, we mainly target high-performance embedded multicore processors, which are commonly 
used in computationally demanding embedded real-time applications such as self-driving cars. 
However, we believe that our cache DoS attacks can potentially be effective in server class 
processors, especially in the context of multi-tenant cloud computing infrastructures 
such as Amazon EC where multiple virtual machines from different users 
may share a physical computing platform. As such, a malicious user may be able to mount a cache DoS 
attack to impact the quality-of-service of the other users of the platform. 
We plan to investigate the effectiveness of cache DoS attacks in server class processors as part of our future work.

\section{Related Work}\label{sec:related}

Micro-architectural denial-of-service (DoS) attacks have been studied for several
different types of shared resources in multicore systems. 
Moscibroda et al. demonstrated DoS attacks on memory (DRAM) controllers~\cite{moscibroda2007memory}.
In particular, they found that the widely used FR-FCFS~\cite{rixner2000memory}
scheduling algorithm, which prioritizes row hits, is susceptible to
DoS attacks. In response, many ``fair'' memory scheduling algorithms (e.g. ~\cite{mutlu2007stall,kim2010thread})
were proposed to balance performance and fairness in scheduling memory.  
Keramidas et al. studied DoS attacks on cache space and proposed a cache
replacement policy that allocates less space to such attackers (or
cache “hungry” threads)~\cite{keramidas2006preventing}. 
Unwanted cache space evictions is particularly well-known type of interference, 
which was well studied by
Woo et al. when they investigated DoS attacks on cache bus (between L1 and L2)
bandwidth, main memory bus (front-side bus) bandwidth, and shared cache
space, on a simulated multicore platform~\cite{woo2007analyzing}.
More recently, our prior works focused on internal
hardware buffers of shared non-blocking caches and demonstrated the
effectiveness and severity of cache DoS attacks~\cite{bechtel2019dos,bechtel2018picar,valsan2016taming}.
Iorga et al. leveraged the DoS attacks from~\cite{bechtel2019dos} and 
presented a statistical testing method to evaluate shared resource interference on a number 
of embedded multicore platforms~\cite{iorga2020slow}. 
The memory-aware cache DoS attacks proposed here are significantly more effective than
prior cache DoS attacks by taking advantage of memory address mapping
information and HugePage support.

Even in the absence of malicious attackers, normal applications sharing a multicore can interfere with each other due to contention on the shared hardware resources, which is especially problematic for real-time systems as they need isolation and timing guarantee. Consequently, there is a large body of work in the real-time systems research community to provide stronger isolation in multicore, most of which have been focused on two major shared resources: shared cache space and main memory bandwidth. Many researchers proposed various software and hardware mechanisms and policies to manage these resources~\cite{mancuso2013rtas,kim2013coordinated,ye2014coloris,kim2017attacking,roozkhosh2020potential,yun2013rtas,ewarp20,farshchi2018deterministic,xu2019holistic}. 
The moves for greater shared resource management has also been seen in industry as well. In fact, major 
CPU manufacturers have added hardware support for shared resource management. For example, some of recent Intel processors include support for the Resource Director Technology (RDT) technology~\cite{intelswref}, which allows for low overhead management of shared cache space and memory bandwidth. ARM also introduced a similar technology called Memory System Resource Partitioning and Monitoring (MPAM)~\cite{arm-mpam-supp}.
Recently, Xu et al., proposed a joint shared cache space and memory bandwidth partitioning technique to provide stronger isolation in multicore~\cite{xu2019holistic}, utilizing both Intel's hardware based cache partitioning and software based memory bandwidth throttling mechanisms.
As we previously showed, however, cache space partitioning techniques do not necessarily protect against cache DoS attacks as they target cache internal buffers which can still be shared even when the cache is partitioned~\cite{bechtel2019dos,valsan2016taming}.
On the other hand, memory bandwidth management (throttling) has been shown to be a viable defense against cache DoS attacks~\cite{bechtel2019dos}. This is because it limits the rate of memory accesses, which is key to any successful DoS attack, including the memory-aware ones we proposed in this work.

\section{Conclusion}\label{sec:conclusion}

In this paper, we introduced memory-aware cache DoS attacks that leverage
a system's memory address mapping information and HugePage support to
induce prolonged cache blocking by intentionally creating DRAM bank
congestion.
From extensive experiments on two popular embedded multicore platforms, we 
show that our memory-aware cache DoS attacks can generate significantly higher timing
impacts to cross-core victim tasks compared to prior cache DoS
attacks. 


\bibliographystyle{abbrv} 
\bibliography{reference}

\end{document}